%% file: sample-manuscript.tex
\documentclass[manuscript,screen]{acmart}

\usepackage{soul}
\usepackage{multirow}
\usepackage{makecell}

\AtBeginDocument{%
  \providecommand\BibTeX{{%
    \normalfont B\kern-0.5em{\scshape i\kern-0.25em b}\kern-0.8em\TeX}}}

\setcopyright{acmcopyright}
\acmYear{Accepted to appear in the CACM}

\begin{document}

\title{An Analysis of the Math Requirements of 199 CS BS/BA Degrees at 158 U.S. Universities}


\author{Carla E.\ Brodley}
\email{c.brodley@northeastern.edu}
\affiliation{%
  \institution{Center for Inclusive Computing at Northeastern University}
  \streetaddress{177 Huntington Ave}
  \city{Boston}
  \state{Massachusetts}
  \country{USA}
  \postcode{02115}
}
\author{McKenna Quam}
\email{quam.m@northeastern.edu}
\affiliation{%
  \institution{Khoury College of Computer Sciences, Northeastern University}
  \streetaddress{177 Huntington Ave}
  \city{Boston}
  \state{Massachusetts}
  \country{USA}
  \postcode{02115}
}

\author{Mark A. Weiss}
\email{weiss@fiu.edu}
\affiliation{%
  \institution{Knight Foundation School of Computing and Information Sciences, Florida International University}
  \streetaddress{11200 SW 8 St}
  \city{Miami}
  \state{Florida}
  \country{USA}
  \postcode{33199}
}

\renewcommand{\shortauthors}{Quam, Brodley, and Weiss}

\begin{abstract}
For at least 40 years, there has been debate and disagreement as to the role of mathematics in the computer science
curriculum. 
This paper presents the results of an analysis of the math requirements of 199 Computer Science BS/BA degrees from 158 U.S. universities, looking not only at which math courses
are required, but how they are used as prerequisites (and corequisites) for computer science (CS) courses.
Our analysis shows that while there is consensus that discrete math is critical for a CS degree, and further that calculus is almost always required for the BS in CS, there is little consensus as to when a student should have mastered these subjects.  Based on our analysis of how math requirements impact access, retention and on-time degree completion for the BS and the BA in CS, we provide several recommendations for CS departments to consider. 
\end{abstract}



\received{January 2024}
\received[Revised]{April 2024}
\received[accepted]{April 2024}

\maketitle

\input{sections/intro}

\input{sections/methodology}

\input{sections/results}

\input{sections/fiu}
\input{sections/related_work}
\input{sections/conclusions}

\section*{Acknowledgments}

This project was partially funded by Pivotal Ventures and by the
National Science Foundation (award 230845). We would like to thank Raj Rajenda, Rahul Simha, Felix Muzny, Monique Ross, Albert Lionelle, Megan Giordano and Catherine Gill for their comments on earlier drafts of this paper.

\bibliographystyle{ACM-Reference-Format}
\bibliography{sample-base}


\end{document}

%% file: sections/intro.tex
\section{Introduction}

This paper presents the results of an analysis of the math requirements of 199 CS BS/BA degrees from 158 U.S. universities.  Our motivation stems from three observations.  First, in prior work on degree complexity  \cite{Lionelle24}, we observed that, in the sixty schools analyzed, mathematics (particularly calculus) represented a barrier to student progression in the CS sequence.  Second, as part of the Center for Inclusive Computing's (CIC)\footnote{The CIC (\url{www.cic.northeastern.edu}) is a national effort to create systemic, sustainable change in U.S. universities to broaden participation in computing.  The CIC works with and funds universities to make systemic changes  to the way in which they offer their introductory CS sequence with the goal that true beginners to computing can discover, thrive and persist \cite{Brodley22CRN, Brodley22CACM, Brodley24CACM, Lionelle24, Muzny23, Muzny24}. Ensuring a pathway for true beginners is important for broadening participation in computing because those who are true beginners are often from populations that have been historically marginalized in CS.} work, we performed all-day site visits at the CS departments of 54 universities and observed significant variation in both the number and placement of required math courses. Third, at Florida International University (FIU) we observed firsthand that a BA in CS, with fewer math requirements than the BS in CS, increased student retention, diversity and demand for computer science. 

The goals of this analysis are to determine: 1) if the variability in the requirements and placement\footnote{By placement we mean how early in the CS curriculum a math course shows up as a pre- or corequisite to a required CS class.} of math classes we observed held for a larger sample of universities; 2) if there are differences in the number of math classes for ABET-accredited\footnote{The Accreditation Board for Engineering and Technology (ABET) \cite{aboutABET} is a non profit organization which certifies engineering/CS programs.} versus non-ABET accredited CS programs and for BA versus BS CS degrees; and 3) commonalities across institutions as to what math is required to be a successful computer scientist.  In this paper we present our methodology, our results, guidelines published by ABET \cite{ABET-CAC} and the ACM \cite{ACM2023gamma}, and we make several recommendations for CS departments to consider.

%% file: sections/methodology.tex
\section{Methodology}

We started with all U.S. post-secondary bachelor degree granting institutions that graduated 150 or more students in CIP\footnote{The Classification of Instructional Programs (CIP) provides a taxonomic scheme that supports the accurate tracking and reporting of fields of study and program completions activity \cite{IPEDS}.  CIP code 11 is for computing majors and includes CS.} code 11 for academic year 2021 as reported in the Integrated Postsecondary Education Data System (IPEDS) \cite{IPEDS}. This led to an initial sample size of 179 distinct universities.  We made the choice to select schools with the largest numbers of graduates to analyze the requirements impacting the majority of students studying CS in the U.S. (for example, if we look at CIP codes 11.0101 and 11.0701 alone -- the two CIP "sub" codes that most commonly contain CS degrees \cite{tims2023-inroads}, our sample size captures 60\% of all undergraduates earning a CS degree in 2021).    

We removed degrees/schools from the sample if 1) the CS major requirements are not publicly available; 2) the degree is interdisciplinary (e.g., Business, or CS+X); and/or 3) the institution is a religious institution with more than 15 credits for religion coursework.\footnote{We were concerned that the focus on religious curriculum would lead the faculty to down select CS and math classes in order to make it feasible for a student to graduate in four years.}   
The resulting sample contained 199 degrees from 158 universities spanning 39 states plus the District of Columbia and including four online universities. The state with the most representation was California with 28 degrees. Of the 158 institutions, 47 are members of the American Association of Universities \cite{AAU}. Further statistics of interest are: the sample contained 80\% BS and 20\% BA degrees; an almost even split between degrees offered by CS departments in and out of engineering (100 and 99 degrees respectively);\footnote{We count degrees from Colleges of Computer Science in the non-engineering category.} 83 of the programs are CAC ABET accredited \cite{ABET-CAC}; and 65 of the degrees are from the 55 Minority Serving Institutions (MSIs) included in the sample. 

For each degree, using  publicly-available degree requirements and plans of study, we recorded all math requirements. In our analysis we made the following design decisions:
\begin{itemize}
    \item For 26 degrees, discrete math is split over two classes;  we counted these as a single class because it did not change the results (e.g., if calculus 1 was required for "Discrete 1" then it is clearly also a prerequisite of "Discrete 2"). Of the 24 degrees with two discrete math classes, six were on the quarter system.   
    \item We defined core CS classes as those that all CS majors must complete.  
    \item We analyze the math requirements only for core CS classes because our focus is on the impact of math requirements on all CS majors at an institution.  Higher-level math courses may be needed for some elective CS classes, but a student can choose to avoid those classes.  
    \item We counted classes labeled as "Data Structures \& Algorithms" twice: once as data structures and once as algorithms.  Schools differ in whether this material is taught over one or two classes because if "Data Structures \& Algorithms"  has a math pre- or corequisite then the math class is blocking the student from both topics. 
    \end{itemize}

%% file: sections/results.tex
\section{Results}
\label{sec:results}

In Table \ref{tab:math_reqs} we report the total number of degrees that require each of the six most prevalent math courses.  For example we see that 191 of the 199 degrees  require calculus 1.   Indeed, we see that calculus 1 and discrete math are required by almost all programs. Calculus 3 is required by only 43 programs (the majority of which are in CS departments housed in a college of engineering and are ABET accredited).  Note that some programs require a  math elective and often calculus 3 is on the list of possible electives.\footnote{Number and choice of math electives is not given in Table \ref{tab:math_reqs}.}

\input{tables/math_reqs}

\input{tables/discrete_counts}

Tables \ref{tab:discrete}, \ref{tab:calc1} and \ref{tab:Calc2} show the pre- and corequisite math classes for core CS classes.   For example, we see in Table \ref{tab:calc1} that calculus 1 is required before discrete math in 64 of the degrees and as a corequisite in nine degrees.  We chose to present the detailed data for calculus 1, calculus 2 and discrete math as these are the three most frequently required math classes and also the ones with the most impact on a student's progression through a program.  More than one core CS class might explicitly list a math class as a prerequisite  (for example a program might explicitly list discrete math as a prerequisite for algorithms and for computer organization because students can take algorithms and computer organization in any order but each can only be taken {\em after} a student completes discrete math). Note that some courses have a math class as a ``transitive prerequisite" but these are not listed in the tables (e.g., if discrete math is a prerequisite for data structures, and data structures is a prerequisite for algorithms, then by transitive closure discrete math is a prerequisite for algorithms).\footnote{We did not calculate the transitive prerequisites as this would have been extremely time-consuming and would not change the key points of this paper.}

We draw six conclusions from the data.  First, 198 of the 199 degrees analyzed require discrete math be taken at some point in the degree.\footnote{The one exception is an  application-development degree.} However, the results shown in Table \ref{tab:discrete} illustrate that there is no national consensus as to which classes should directly require discrete and whether it should be listed as a pre- or corequisite.  

Second, 81\% of the BS degrees require both calculus 1 and calculus 2, whereas only 57\% of the BA degrees require both. {\em When} calculus is required varies substantially \cite{Lionelle24}.  This matters because the prescribed timing of when a student must take calculus can have a large impact on time-to-degree. At one end of the spectrum, 33 degree plans have calculus 1 as a pre- or corequisite to CS 1, meaning that a student has to be "calculus ready" in 24 degrees and have completed calculus 1 in 13 degrees to even begin their CS journey. In many universities a substantial proportion of incoming first year students are not calculus-ready and thus their discovery of CS is barred by this requirement (note that only about 20\% of high school students take calculus \cite{Bressoud-2021}). A question that arises from this variability is whether the requirement of calculus as an early prerequisite stems from an actual need in CS courses or whether it is a historical artifact of how CS programs were constructed decades ago (e.g., most CS departments grew out of mathematics or computer engineering departments).

Third, there are 64 degrees that require that  calculus 1 be completed before discrete math and 13 that require calculus 2 before a student is allowed to take discrete math.  This also has major implications for time-to-degree because if a student fails calculus then they are delayed in taking discrete math, which along with data structures are the two courses that almost all higher-level CS classes require \cite{Lionelle24}. Because many schools have high drop/fail/withdraw (DFW) rates for calculus (in the CIC's portfolio of Implementation Grantees\footnote{See cic.northeastern.edu for a list of all Implementation Grantees.} some rates are as high as 50\%), this requirement can easily add one to two semesters to a student's time-to-degree.

\input{tables/calc_1_counts}
\input{tables/calc_2_counts}

Fourth, which department teaches discrete math also varies.  Discrete is taught by the mathematics department in 51 degrees, by the CS department in 130 degrees, and for 17 degrees a student can take discrete math from either department.  We examined whether the prerequisites for discrete math differ depending on the offering department.    Of the 51 degrees for which discrete math is taught by the math department, 19 have calculus as a prerequisite and for the 130 degrees for which discrete is taught by CS, 61 degrees require calculus as a prerequisite.  

\begin{figure}[htp]
    \centering
    \includegraphics[width=7cm]{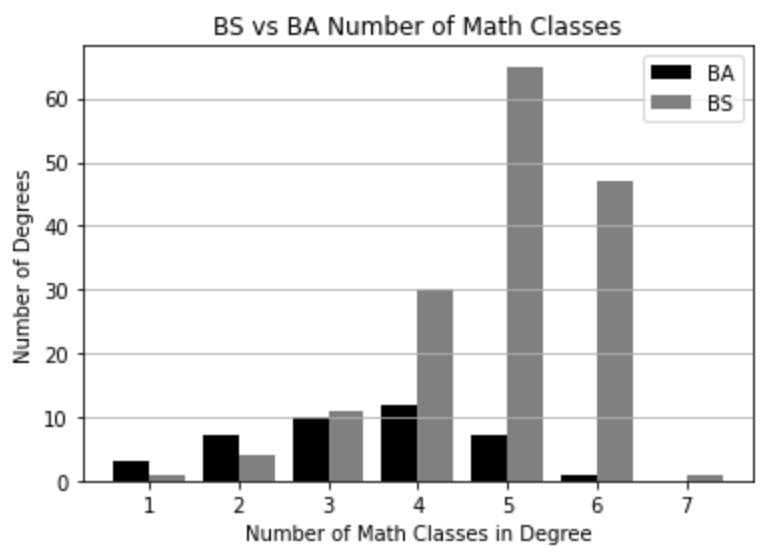}
    \caption{Distribution of total math classes required for BA versus BS degrees in the sample.}
    \label{fig:BA-BS-number}
\end{figure}

\begin{figure}[htp]
    \centering
    \includegraphics[width=7cm]{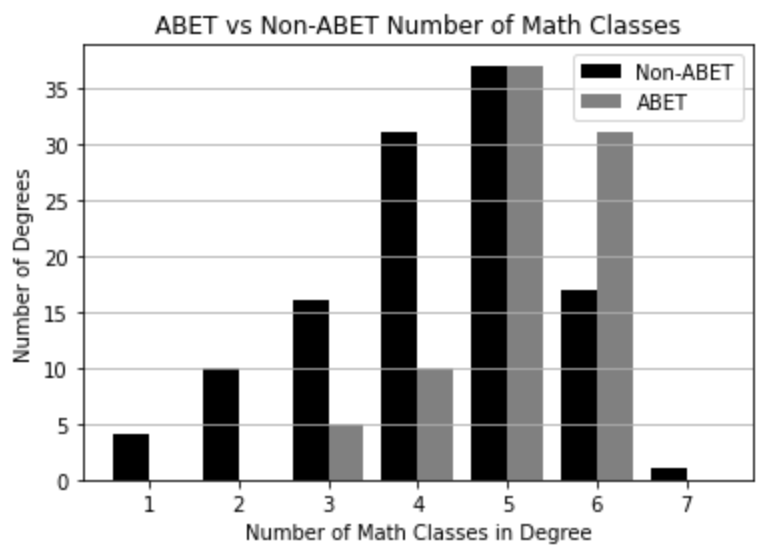}
    \caption{Distribution of total math classes required for ABET accredited and non ABET accredited degrees in the sample.}
    \label{fig:ABET-number}
\end{figure}

Fifth, not surprisingly, there are significant differences as to the number and placement of math classes for BA and BS degrees. Calculus 1 and 2 are more frequently pre- and corequisites to CS1, CS2, and discrete math in BS degrees than in BA degrees (e.g.,  calculus 1 is a prerequisite for discrete math in 41.5\% of the BS degrees  compared to just  27.5\% in the BA degrees). Furthermore, BA degrees require fewer math classes than BS degrees, with an average of 3.4 and 4.9 required math classes, respectively.  The distribution of math requirements in BA and BS degrees is shown in  Figure \ref{fig:BA-BS-number}. 

Finally, we examined the difference in math requirements for ABET and non-ABET accredited programs. On average an ABET accredited program requires 5.1 math classes and a non-ABET accredited program requires 4.2 math classes, and this difference is statistically significant.  We show the distribution in Figure \ref{fig:ABET-number}. Note that for BS degrees that are not ABET accredited the average number of math classes is 4.6. Although ABET requires 15 credits of math for CS, schools vary in the number of credits assigned to each class and some count ``highly-mathematical" CS classes toward this requirement.\footnote{This is permitted as long as they have different courses in CS to meet the computing credits requirement. That is, programs cannot double count CS courses as Math and as CS to meet the ABET math and CS credit requirements.}

%% file: tables/math_reqs.tex
\begin{table}[]
    \centering
    \begin{tabular}{|l|r|r|}
         \hline
         Course & Degrees & Percentage\\
         \hline
         Calculus 1 & 191 & 96.0 \\
         \hline
         Calculus 2 & 152 & 76.4 \\
         \hline
         Discrete & 198 & 99.5 \\
         \hline
         Probability/Statistics & 140 & 70.4 \\
         \hline
         Linear Algebra &  116 & 58.3 \\
         \hline
         Calculus 3 & 43 & 21.6 \\
         \hline
    \end{tabular}
    \caption{For each math class we list the number and percentage of the 199 degrees for which the class is required.}
    \label{tab:math_reqs}
\end{table}

%% file: tables/discrete_counts.tex
\begin{table}[]
    \centering

\begin{tabular}{|p{4 cm}|r|r|r|}
     \hline
     Courses for which Discrete is a & Pre-req & Co-req & Total \\
     \Xhline{3\arrayrulewidth}
     Algorithms & 115 & 2 & 117 \\
     \hline
     Data Structures & 69 & 4 & 73 \\
     \hline
     Theory of Computation & 28 & & 28 \\
     \hline
     Programming Languages & 22 & & 22 \\
     \hline
     Computer Organization & 18 & & 18 \\
     \hline
     Intro to Databases & 17 & & 17 \\
     \hline
     Computer Systems & 12 & & 12 \\
     \hline
     Automata & 12 & & 12 \\
     \hline
     Computer Architecture & 11 & & 11 \\
     \hline
     Formal Lang/Assembly & 9 & & 9 \\
     \hline
     Software Dev/Engr & 8 & 1 & 9 \\
     \hline
     CS 2 & 6 & 3 & 9 \\
     \hline
     Artificial Intelligence & 4 & & 4 \\
     \hline
     Operating Systems & 4 & & 4 \\
     \hline
     Object Oriented Design & 4 & & 4 \\
     \hline
     Intro to Cybersecurity & 3 & & 3 \\
     \hline
     CS 3 & 3 & & 3 \\
     \hline
     Intro to Data Science & 3 & & 3 \\
     \hline
     Machine Learning & 3 & & 3 \\
     \hline
     Computer Networks & 3 & & 3 \\
     \hline
     Randomness \& Composition & 2 & & 2 \\
     \hline
     Digital Logic & 2 & & 2 \\
     \hline
     Great Ideas in CS & 1 & & 1 \\
     \hline
     Intro to CS and Engineering & 1 & & 1 \\
     \hline
     Cryptography & 1 & & 1 \\
     \hline
     Secure Programming & 1 & & 1 \\
     \hline
     Parallel Computing & 1 & & 1 \\
     \hline
     CS 1 & & 1 & 1 \\
     \Xhline{3\arrayrulewidth}
     Total & 363 & 11 & 374 \\
     \hline
     No Post/Co-Req & & & 18 \\
     \hline 
\end{tabular}
    \caption{Courses for which Discrete Math is a direct pre-requisite or co-requisite (listed in descending order).}
    \label{tab:discrete}
\end{table}

%% file: tables/calc_1_counts.tex
\begin{table}[]
    \centering
    
\begin{tabular}{|p{4.5 cm}|r|r|r|}
     \hline
     Courses for which Calculus 1 is a  & Pre-req & Co-req & Total \\
     \Xhline{3\arrayrulewidth}
     Discrete &  64 & 9 & 73 \\
     \hline
     CS 1 & 13 & 24 & 37 \\
     \hline
     CS 2 & 12 & 11 & 23 \\
     \hline
     Data Structures & 11 & 5 & 16 \\
     \hline
     Algorithms & 8 & & 8 \\
     \hline
     Software Dev/Engr & 6 & & 6 \\
     \hline
     Digital Logic & 2 & 1 & 3 \\
     \hline
     Computer Architecture & 2 & & 2 \\
     \hline
     Computer Organization & 2 & & 2 \\
     \hline
     Theory of Computation & 2 & & 2 \\
     \hline
     CS 3 & 1 & 1 &  2 \\
     \hline
     Computer Systems & 1 & & 1\\
     \hline
     Operating Systems & 1 & & 1 \\
     \hline
     Numerical Methods & 1 & & 1 \\
     \hline
     Parallel Computing & 1 & & 1 \\
     \hline
     Computer Networks & & 1 & 1 \\
     \hline
     Machine Learning & & 1 & 1 \\
     \Xhline{3\arrayrulewidth}
     Total & 128 & 54 & 180 \\
     \hline
     No Post/Co Req & & & 75 \\
     \hline
\end{tabular} 
    \caption{Courses for which Calculus 1 is a direct pre-requisite or co-requisite (listed in descending order).} 
    \label{tab:calc1}
\end{table}

%% file: tables/calc_2_counts.tex
\begin{table}[]
    \centering
    
\begin{tabular}{|p{4.5 cm}|r|r|r|}
     \hline
     Courses for which Calculus 2 is a  & Pre-req & Co-req & Total \\
     \Xhline{3\arrayrulewidth}
     Discrete & 13 & 2 & 15 \\
     \hline
     Algorithms & 14 & & 14 \\
     \hline
     Data Structures & 4 & 1 & 5 \\
     \hline
     Computer Systems & 1 & 2 & 3 \\
     \hline
     Theory of Computation & 2 & & 2 \\
     \hline
     CS 2 & & 2 & 2 \\
     \hline
     CS 4 & 1 & & 1 \\
     \hline
     Operating Systems & 1 & & 1 \\
     \hline
     Introduction to Data Science & 1 & & 1 \\
     \hline
     Numerical Methods & 1 & & 1 \\
     \Xhline{3\arrayrulewidth}
     Total & 38 & 7 & 45 \\
     \hline
     No Post/Co Req & & & 108 \\
     \hline
\end{tabular}

     \caption{Courses for which Calculus 2 is a direct pre-requisite or co-requisite (listed in descending order).} 
    \label{tab:Calc2}
\end{table}

%% file: sections/fiu.tex
\section{A Case Study of the Impact of Math in the CS Curriculum}
\label{sec:FIU}

At Florida International University (FIU), the considerations cited in this paper (timing of math classes, linkages between math classes and the CS sequence, etc) came into play when they examined their ABET accredited BS in CS program in 2015. FIU is a large, urban, minority-serving public university with a majority of "non-traditional students," which the National Center for Education Statistics defines as students who either delay enrollment, attend part time, work full time, are financially independent for purposes of determining eligibility for financial aid, have dependents other than a spouse, are a single parent, and/or do not have a high school diploma \cite{nontraditional}. Furthermore, the majority of FIU undergraduates receive Pell grants, meaning they are considered low-income. For these students, graduating with little to no debt into a well-paying job can be life changing.


At FIU, the CS department is housed in the engineering college. In 2015, like most programs, FIU's CAC ABET accredited program required calculus 1 and 2 (both four-credit courses), discrete math,\footnote{During this period, FIU's computer science faculty created their own discrete structures course as an alternative to the math department's offering, allowing students to choose (and knowing their likely choice).  This change increased retention and progression.}  statistics, and automata theory, a course which allowed the program to meet the ABET math requirement. In addition, like the rest of engineering, FIU required physics 1 and 2 with labs (five credits each) in addition to other science courses. In other words, the CS major was very math and science heavy compared to other non-engineering degrees. Eliminating ABET accreditation was not an option, but the department was interested in offering a less math and science heavy CS degree as many students were either not successful in completing the BS in CS degree or were deciding to not pursue the degree because of the added time to degree completion and/or perceived difficulty.

In FIU's CS curriculum, calculus 1 and 2 were not prerequisites for any required CS course, but were required for physics  and statistics.  Many students, knowing the importance of completing certain CS courses to be able to apply for  internships, and wanting to avoid semesters that included physics, calculus, and programming courses, opted to load up on CS courses early in their studies and delay the  "math+physics" part. It was not uncommon to have students complete all the CS course work successfully, have a good job offer, but then have graduation held-up because they did not pass one or more of statistics, physics 2,  calculus 2, or automata theory in their final semester. 


As the university focused increasingly on reducing time to graduation, students were denied opportunities to switch majors after the first 1-2 years because they could not possibly do so without significant delay in graduation time. Additionally, FIU observed an unfortunate reality that many transfer students self-selected not to be CS majors, even if that had been their intent when starting at FIU. Although Florida universities in theory have strong articulation agreements with their state colleges, in practice many transfer students who were interested in CS found that they didn't have the right math courses to graduate on time.  In 2015, the calculus and physics requirements in the CS degree added 18 credits to what was an already packed 60 upper division credits.  This also added significantly more cost to the degree as Florida universities are mandated to charge double tuition for "excess hours" \cite{excesshours}. Because FIU's state college transfer student population (those coming to FIU with an Associate's degree from a two-year institution) represent more than half of the university's students, this meant large numbers of students were not pursuing computer science for reasons that were not tied to the core computer science curriculum.  Another consideration was that, even pre-COVID, many working students wanted the flexibility of at least some online courses and the math and physics courses at FIU had limited, to no offerings, online because of poor student success in that modality.

With all this, FIU decided to keep its ABET accredited BS in CS and create a new BA in CS, distinguished primarily by 1) eliminating the calculus and physics requirements; 2) replacing the statistics course with one of the math department alternatives that does not require calculus as a prerequisite; and 3) moving a few required courses (such as automata theory) into the elective group.  By requiring only discrete structures and statistics, the math and science requirement for the BA in CS  became only slightly more than  what is required for students in other non-CS and engineering majors. To be clear, although the BS and BA differ in the number of required versus elective CS courses, the main reason students cite in distinguishing the two programs is whether the calculus sequence is required. 


FIU's BA in CS alternative has been wildly popular.  Today, the BA degree accounts for more than 60\% of CS majors, but in absolute numbers the BS program has also grown. Above all, students are empowered to choose the pathway that best meets their needs. Many BA students indicate they would have not been able to start the CS degree without this pathway. Importantly, the graduation rates in CS, which lagged the rest of the university, have caught up, alleviating the perception among students that the CS major was not worth pursuing.

Preliminary results presented in \cite{ross2022removing}  indicated that while there was an immediate significant increase in both the number of computing majors overall and in the percentage of women pursing CS degrees, there were no statistically significant differences between the BA and BS degrees in enrollment on the basis of gender or on job attainment. There was a statistically  significant difference in enrollment on the basis of race/ethnicity; a larger percentage of Hispanic, Asian, and Nonresident Alien students were enrolled in the BS program compared with other ethnic groups, while the BA program attracted a much higher percentage of Black students than the BS. Surveys completed by 45 BA and 89 BS graduates showed no statistically significant difference in obtaining job offers, but a statistical difference in salary distribution, favoring the BS degree. Beyond the relatively small numbers of responses to the employment survey, a notable limitation of the study was that at the time of the survey distribution, students were job hunting in the height of COVID.

With the exception of the salary disparity, many of the preliminary results have been confirmed in a more recent analysis. Notably, the percentage of computing majors who identify as Black or African American has risen from 11.0\% to 13.0\% since the full launch of the BA degree, even as the corresponding percentage has declined from roughly 12.5\% to 11.3\% of the overall FIU undergraduate student body. More recent survey results also show that the salary disparity between BA and BS pathways has disappeared. Details on this work, including results from a qualitative mixed-methods analysis are forthcoming \cite{ross2024}.


Of course every institution is unique and has its own circumstances, but FIU's example shows that changes to required courses, prerequisites and corequisites can have significant impact.

%% file: sections/related_work.tex
\section{Related Work}
\label{related work}


In this section we review the existing literature on the role of mathematics in CS and the recommendations of the two professional organizations that provide curricular recommendations and guidelines for CS.  The first,  the Accreditation Board for Engineering and Technology (ABET) \cite{aboutABET}, was founded in 1932 to provide guidelines and oversight of Engineering programs.  CS accreditation started in 1985 under the Computer Science Accreditation Board (CSAB), and is currently conducted by ABET's Computing Accreditation Commission \cite{ABET-CAC}.  The second, the Association for Computing Machinery (ACM), was founded in 1947 and is a U.S.-based international professional society for computing \cite{ACM-general-site}.  ACM publishes curricular guidelines for CS every ten years with the two most recent publications being ACM-2023-Gamma \cite{ACM2023gamma} and ACM-2013 guidelines \cite{ACM2013}.  

\subsection{ABET-CAC Guidlines}

ABET is a non-profit organization that certifies engineering/computing programs (as well as applied and natural science and engineering technology).  Its goal is to provide employers and students a guarantee of what knowledge and skills graduates will have mastered.  Although likely all engineering colleges require their engineering programs to be ABET accredited, this is not true of CS programs even when the CS department sits within the college of engineering.   Indeed only 83 of the 199 programs studied in our sample are CAC ABET accredited.

The 2024-2025 CAC ABET accreditation requirements state that for a CS program to be accredited it must contain 15 credit hours of mathematics and statistics, which must include discrete mathematics, probability, statistics, and have rigor equivalent to introductory calculus \cite{ABET-CAC}. Note that calculus 1 is not explicitly required. Rather, the wording "rigor equivalent to introductory calculus," which was first introduced in the 2019 criteria, only precludes courses such as pre-calculus from counting toward the 15 credit hours \cite{Oudshoorn18}: "the intent is to signal to programs that calculus is not necessarily required if it does not support the program’s objectives as well as other mathematics courses...  it is hoped that programs see it as an opportunity to include mathematics courses that best support the program rather than simply requiring traditional courses such as calculus." Despite this, none of the CAC ABET accredited programs in our sample have removed calculus 1 in response to this change and only a handful have removed calculus 2. Further, the accreditation requirements make no mention of how classes should be ordered. The report does not specify which CS topics require particular math concepts and thus is silent on which CS courses should require calculus (or any other math class) as a prerequisite.  Some CAC ABET accredited CS programs code computational complexity and/or algorithms as math courses, which is permitted assuming they have a sufficiently mathematical focus. Also of note, ABET does not require that the credit hours be accounted for in complete courses. So, for instance, in a program that requires calculus 1 and 2, discrete math, and statistics, but for which calculus 1 and 2 are both four credits, the 15-credit hour requirement can be met by finding one credit somewhere inside a three-credit CS course, and counting that course as one math credit and two CS credits, rather than finding a fifth math course.

Although not a subject of this paper, it is worth noting that that this year ABET CAC replaced the previous requirement of six credits of science, with the following requirement: "Coursework that develops and applies the scientific method in a non-computing area" \cite{ABET-CAC}. Thus just {\it a single science class which meets these criteria is sufficient for CAC ABET 2024 accreditation}, which, for most universities, is no more than the general education requirements for all majors.  Schools that find themselves struggling to fit all of the CS classes into their degrees as the discipline continues to evolve and expand would do well by re-examining their legacy science requirements as a first step.

\subsection{ACM 2023 Guidelines}

The ACM 2023 Computer Science Curricula committee issues curricular guidelines for CS undergraduate programs.  This is a partnership of ACM, IEEE-CS and AAAI \cite{ACM2023gamma}.  The current version is still in draft format, but has been published to ACM's website (\url{https://csed.acm.org}). 

Historically, the ACM curricular guidelines have {\em not} recommended that students take calculus before CS1 or discrete math \cite{ACM2013,ACM68,ACM78}.  In the {\em Math Requirements} section, the 2023 report recommends that students should come to college with pre-calculus, and thus pre-calculus can be viewed as a prerequisite for discrete math. The {\em Core Hours} section outlines two sets of core math requirements: a minimal "CS-core" set suited to credit-limited majors and a more expansive "KA-core" set to align with technically focused programs.  The only course needed to satisfy the "CS-core" set is discrete math.  The "KA-core" set contains some coverage of calculus, probability and linear algebra in addition to discrete math, but does not specify  how many credit hours should be devoted to their study. The {\em Course Packing Suggestions} section reports: "Many computer science departments
now offer courses that prepare students mathematically for AI and machine learning. Such courses can combine just enough calculus, optimization, linear algebra and probability; yet others may split
linear algebra into its own course. These courses have the advantage of motivating students with
computing applications, and including programming as pedagogy for mathematical concepts." These CS departments have created  their own math curriculum so that they are not forced to require CS students take the same math curriculum used in engineering and mathematics departments.  This is the approach taken at Northeastern University where all CS majors are required to take discrete math, calculus 1 and CS2810 Mathematics of Data Models.\footnote{CS 2810
studies the methods and ideas in linear algebra, multivariable calculus, and statistics that are most relevant for the practicing computer scientist doing machine learning, modeling, or hypothesis testing with data. The course covers least squares regression, finding eigenvalues to predict a linear system’s behavior, performing gradient descent to fit a model to data, and performing t-tests and chi-square tests to determine whether differences between populations are significant. The course includes applications to popular machine-learning methods, including Bayesian models and neural networks.} and at George Washington University which offers CS-4341/6341: Continuous Structures (https://www2.seas.gwu.edu/~simhaweb/contalg/instructors.html). The ACM's flexibility around math is recent, as, historically, the ACM curricular guidelines have favored highly technical math. Indeed, M. Tedre's history of CS education points out that the 1968 ACM guidelines were written by mathematicians not computer scientists \cite{histCS}. 

\subsection{Research on Mathematics Requirements}

For at least 40 years, there has been debate and disagreement as to the role of mathematics in the computer science curriculum. Some question whether the historically high amount of math in many CS programs is still necessary \cite{dougherty2017math, ralston1981computer, ralston2005we}, some suggesting that the fundamental math course for CS is discrete math \cite{bruce03}, and others arguing that the community has become math-phobic \cite{beaubouef02,tucker2001our}. The overarching argument is that mathematics proficiency often correlates with success in computing \cite{konvalina1983math,wilson2001contributing} and that mathematics reasoning and problem solving translate to skills needed for CS \cite{baldwin2013}. But the evidence is less clear.
For instance, studies twenty years ago had mixed results as to  whether math is a predictor of success in programming \cite{bennedsen05,ventura05}. More recent work \cite{chen2020high} suggests that taking AP Calculus in high school produced better outcomes in introductory programming courses than taking AP CS.  However, when controlling for students' background characteristics (such as parental education),  that advantage disappeared. Among all this confusion, there are studies suggesting that mathematics requirements in STEM programs, including engineering and computing  serve as a barrier to broadening participation, weeding out students who could otherwise succeed \cite{cheema2013analyzing,Ellis16}.  Indeed Bressoud's seminal 2015 study provides a thoughtful discussion of how the current state of college calculus can deter students from mathematics and STEM majors \cite{Bressoud-2015}. 

To the best of our knowledge, this paper presents the first large-scale analysis in the last decade of the math pre- and corequisites listed on the publicly available websites for universities offering CS degrees in the U.S.\footnote{The ACM 2023 committee made their recommendations from the following inputs: 600 responses of a survey distributed to computer science faculty across a variety of institutional types and in various countries; 865 responses from a survey sent to members of the tech industry; the  math requirements stated by all knowledge areas in the report; from direct input provided by the CS theory community; and based on a review of past ACM reports including recent reports on data science and quantum computing education.}  Papers exist on the design of the CS undergraduate curriculum \cite{berztiss1987mathematically} and on the relationship between math and CS with arguments for CS being reliant on math \cite{LOCKWOOD2019100688}, and for math being reliant on CS \cite{csForProb,de2015some}.  Perhaps the paper most relevant to this work is Baldwin, et al's \cite{baldwin2013} article mapping the connections between math and CS as laid out in the 2013 ACM guidelines \cite{ACM2013}.  They found that the most connected topic to CS was probability and statistics with content connections to higher-level CS topics such as theory of computation, information systems, networks, and computer organization. In their paper, Baldwin, et al presented a survey of 25 degree programs, for which they found the average number of required math classes is 4.5 -- similar to the results of this paper.  

%% file: sections/conclusions.tex
\section{Conclusions and Recommendations}

 Our analysis shows that although the majority of CS departments in the U.S.\ require calculus, there is no consensus as to where in the degree it should be placed. And further that the subject material covered in calculus is likely not needed for most CS classes.  It is our belief that the worst placement is as a pre- or corequisite for CS1 as this impacts a student's ability to discover CS.  Indeed, as stated, many students are not calculus-ready when they enter college and thus are at risk of being delayed in taking CS1 or, worse, dissuaded from trying out and discovering the field. Only half of high schools in the U.S. offer CS so we absolutely must have a good path for students to discover CS at university. Indeed, the degree plans for 162 of the 199 CS programs share the belief that students can take CS1 without calculus as a pre- or corequisite.  Finally, even when it is only listed as a co-requisite, it is not clear that taking calculus 1 and CS1 in the same semester leads to success, particularly for students who are true beginners to coding.

Perhaps the largest variation across the universities is the placement of discrete math in the program.  We conjecture that this is because CS as a field does not yet agree on what topics need be included in discrete math, which in turn causes variation not only in the prerequisites for discrete math but also the post-requistites (i.e., which core classes require discrete math). We know anecdotally that some faculty use math prerequisites for CS classes, not because the subject material of the math course is needed, but because they want a certain "mathematical maturity" before students take the class.  But what does mathematical maturity mean? If the content of a CS course does not rely on the content of a math course but instead the quantitative skills a student would build by taking that math course, then perhaps they can develop those skills in the CS course itself? 

Through this analysis we arrive at two recommendations for CS programs.   First,
do not require calculus 1 or 2 as a prerequisite or co-requisite for CS1, CS2, or discrete math. The first class in which these math concepts {\em might} be used is likely data structures or algorithms (i.e., when  students start to analyze run time and use big-O notation).  Second, CS departments should separate progression in CS from progression in mathematics.  This is particularly important if a university has a high DFW rate in calculus 1 and/or calculus 2.  In this case, calculus blocks many students from being able to move forward in their CS classes. Indeed we recommend that every CS department look at the DFW rates of calculus 1 and  calculus 2 at their university and strongly consider whether these classes are serving their students at all.  And this is particularly important for departments with capacity constraints on the CS major, which use students' GPA in CS1, CS2 and calculus to determine who can major in CS \cite{Brodley22CACM}. Third, as mentioned in the ACM 2023 guidelines, departments might want to consider creating a math class taught by their own faculty to teach the CS-relevant topics from continuous math.    

In summary, we invite you to ponder this ultimately puzzling question:  Are we letting the math department determine who majors in CS?

%% file: sample-manuscript.bbl

\begin{thebibliography}{42}


\ifx \showCODEN    \undefined \def \showCODEN     #1{\unskip}     \fi
\ifx \showDOI      \undefined \def \showDOI       #1{#1}\fi
\ifx \showISBNx    \undefined \def \showISBNx     #1{\unskip}     \fi
\ifx \showISBNxiii \undefined \def \showISBNxiii  #1{\unskip}     \fi
\ifx \showISSN     \undefined \def \showISSN      #1{\unskip}     \fi
\ifx \showLCCN     \undefined \def \showLCCN      #1{\unskip}     \fi
\ifx \shownote     \undefined \def \shownote      #1{#1}          \fi
\ifx \showarticletitle \undefined \def \showarticletitle #1{#1}   \fi
\ifx \showURL      \undefined \def \showURL       {\relax}        \fi
\providecommand\bibfield[2]{#2}
\providecommand\bibinfo[2]{#2}
\providecommand\natexlab[1]{#1}
\providecommand\showeprint[2][]{arXiv:#2}

\bibitem[AAU(2023)]%
        {AAU}
\bibfield{author}{\bibinfo{person}{AAU}.} \bibinfo{year}{2023}\natexlab{}.
\newblock \bibinfo{title}{Who We Are}.
\newblock
\newblock
\urldef\tempurl%
\url{https://www.aau.edu/who-we-are}
\showURL{%
\tempurl}
\newblock
\shownote{(Retrieved December 2023)}.


\bibitem[ABET(2023a)]%
        {aboutABET}
\bibfield{author}{\bibinfo{person}{ABET}.} \bibinfo{year}{2023}\natexlab{a}.
\newblock \bibinfo{title}{About ABET}.
\newblock
\newblock
\urldef\tempurl%
\url{https://www.abet.org/about-abet/}
\showURL{%
\tempurl}
\newblock
\shownote{(Retrieved December 2023)}.


\bibitem[ABET(2023b)]%
        {ABET-CAC}
\bibfield{author}{\bibinfo{person}{ABET}.} \bibinfo{year}{2023}\natexlab{b}.
\newblock \bibinfo{title}{Criteria for Accrediting Computing Programs 2024-2025}.
\newblock \bibinfo{howpublished}{\url{https://www.abet.org/accreditation/accreditation-criteria/criteria-for-accrediting-computing-programs-2024-2025/}}.
\newblock
\newblock
\shownote{(Retrieved December 2023)}.


\bibitem[ACM(2023)]%
        {ACM-general-site}
\bibfield{author}{\bibinfo{person}{ACM}.} \bibinfo{year}{2023}\natexlab{}.
\newblock \bibinfo{title}{ACM Homepage}.
\newblock
\newblock
\urldef\tempurl%
\url{https://www.acm.org/}
\showURL{%
\tempurl}
\newblock
\shownote{(Retrieved December 2023)}.


\bibitem[Atchison et~al\mbox{.}(1968)]%
        {ACM68}
\bibfield{author}{\bibinfo{person}{William~F. Atchison}, \bibinfo{person}{Samuel~D. Conte}, \bibinfo{person}{John~W. Hamblen}, \bibinfo{person}{Thomas~E. Hull}, \bibinfo{person}{Thomas~A. Keenan}, \bibinfo{person}{William~B. Kehl}, \bibinfo{person}{Edward~J. McCluskey}, \bibinfo{person}{Silvio~O. Navarro}, \bibinfo{person}{Werner~C. Rheinboldt}, \bibinfo{person}{Earl~J. Schweppe}, {et~al\mbox{.}}} \bibinfo{year}{1968}\natexlab{}.
\newblock \showarticletitle{Curriculum 68: Recommendations for academic programs in computer science: A report of the ACM curriculum committee on computer science}.
\newblock \bibinfo{journal}{\emph{Commun. ACM}} \bibinfo{volume}{11}, \bibinfo{number}{3} (\bibinfo{year}{1968}), \bibinfo{pages}{151--197}.
\newblock


\bibitem[Austing et~al\mbox{.}(1979)]%
        {ACM78}
\bibfield{author}{\bibinfo{person}{Richard~H. Austing}, \bibinfo{person}{Bruce~H. Barnes}, \bibinfo{person}{Della~T. Bonnette}, \bibinfo{person}{Gerald~L. Engel}, {and} \bibinfo{person}{Gordon Stokes}.} \bibinfo{year}{1979}\natexlab{}.
\newblock \showarticletitle{Curriculum'78: {R}ecommendations for the undergraduate program in computer science—A report of the ACM curriculum committee on computer science}.
\newblock \bibinfo{journal}{\emph{Commun. ACM}} \bibinfo{volume}{22}, \bibinfo{number}{3} (\bibinfo{year}{1979}), \bibinfo{pages}{147--166}.
\newblock


\bibitem[Baldwin et~al\mbox{.}(2013)]%
        {baldwin2013}
\bibfield{author}{\bibinfo{person}{Douglas Baldwin}, \bibinfo{person}{Henry~M. Walker}, {and} \bibinfo{person}{Peter~B. Henderson}.} \bibinfo{year}{2013}\natexlab{}.
\newblock \showarticletitle{The roles of mathematics in computer science}.
\newblock \bibinfo{journal}{\emph{ACM Inroads}} \bibinfo{volume}{4}, \bibinfo{number}{4} (\bibinfo{date}{dec} \bibinfo{year}{2013}), \bibinfo{pages}{74–80}.
\newblock
\showISSN{2153-2184}
\urldef\tempurl%
\url{https://doi.org/10.1145/2537753.2537777}
\showDOI{\tempurl}


\bibitem[Bar‐On and Or‐Bach(1988)]%
        {csForProb}
\bibfield{author}{\bibinfo{person}{Ehud Bar‐On} {and} \bibinfo{person}{Rachel Or‐Bach}.} \bibinfo{year}{1988}\natexlab{}.
\newblock \showarticletitle{Programming mathematics: {A} new approach in introducing probability to less able pupils}.
\newblock \bibinfo{journal}{\emph{International Journal of Mathematical Education in Science and Technology}} \bibinfo{volume}{19}, \bibinfo{number}{2} (\bibinfo{year}{1988}), \bibinfo{pages}{281--297}.
\newblock
\urldef\tempurl%
\url{https://doi.org/10.1080/0020739880190207}
\showDOI{\tempurl}
\showeprint{https://doi.org/10.1080/0020739880190207}


\bibitem[Beaubouef(2002)]%
        {beaubouef02}
\bibfield{author}{\bibinfo{person}{Theresa Beaubouef}.} \bibinfo{year}{2002}\natexlab{}.
\newblock \showarticletitle{Why computer science students need math}.
\newblock \bibinfo{journal}{\emph{ACM SIGCSE Bulletin}} \bibinfo{volume}{34}, \bibinfo{number}{4} (\bibinfo{date}{dec} \bibinfo{year}{2002}), \bibinfo{pages}{57–59}.
\newblock
\showISSN{0097-8418}
\urldef\tempurl%
\url{https://doi.org/10.1145/820127.820166}
\showDOI{\tempurl}


\bibitem[Bennedsen and Caspersen(2005)]%
        {bennedsen05}
\bibfield{author}{\bibinfo{person}{Jens Bennedsen} {and} \bibinfo{person}{Michael~E. Caspersen}.} \bibinfo{year}{2005}\natexlab{}.
\newblock \showarticletitle{An investigation of potential success factors for an introductory model-driven programming course}. In \bibinfo{booktitle}{\emph{Proceedings of the First International Workshop on Computing Education Research}} (Seattle, WA, USA) \emph{(\bibinfo{series}{ICER '05})}. \bibinfo{publisher}{Association for Computing Machinery}, \bibinfo{address}{New York, NY, USA}, \bibinfo{pages}{155–163}.
\newblock
\showISBNx{1595930434}
\urldef\tempurl%
\url{https://doi.org/10.1145/1089786.1089801}
\showDOI{\tempurl}


\bibitem[Berztiss(1987)]%
        {berztiss1987mathematically}
\bibfield{author}{\bibinfo{person}{Alfs Berztiss}.} \bibinfo{year}{1987}\natexlab{}.
\newblock \showarticletitle{A mathematically focused curriculum for computer science}.
\newblock \bibinfo{journal}{\emph{Commun. ACM}} \bibinfo{volume}{30}, \bibinfo{number}{5} (\bibinfo{year}{1987}), \bibinfo{pages}{356--365}.
\newblock


\bibitem[Bressoud(2015)]%
        {Bressoud-2015}
\bibfield{author}{\bibinfo{person}{David Bressoud}.} \bibinfo{year}{2015}\natexlab{}.
\newblock \showarticletitle{Insights from the MAA National Study of College Calculus}.
\newblock \bibinfo{journal}{\emph{The Mathematics Teacher}}  \bibinfo{volume}{109} (\bibinfo{date}{10} \bibinfo{year}{2015}), \bibinfo{pages}{178}.
\newblock
\urldef\tempurl%
\url{https://doi.org/10.5951/mathteacher.109.3.0178}
\showDOI{\tempurl}


\bibitem[Bressoud(2021)]%
        {Bressoud-2021}
\bibfield{author}{\bibinfo{person}{David~M.\ Bressoud}.} \bibinfo{year}{2021}\natexlab{}.
\newblock \showarticletitle{The strange role of calculus in the United States}.
\newblock \bibinfo{journal}{\emph{ZDM: Mathematics Education}} \bibinfo{volume}{53}, \bibinfo{number}{3} (\bibinfo{date}{Jun} \bibinfo{year}{2021}), \bibinfo{pages}{521--533}.
\newblock


\bibitem[Brodley(2022a)]%
        {Brodley22CRN}
\bibfield{author}{\bibinfo{person}{Carla~E. Brodley}.} \bibinfo{year}{2022}\natexlab{a}.
\newblock \showarticletitle{Expanding the pipeline: Addressing the distribution of prior experience in CS1}.
\newblock \bibinfo{journal}{\emph{Computing Research News}} \bibinfo{volume}{34}, \bibinfo{number}{6} (\bibinfo{date}{June} \bibinfo{year}{2022}).
\newblock
\urldef\tempurl%
\url{http://doi.acm.org/10.1145/1219092.1219093}
\showURL{%
\tempurl}


\bibitem[Brodley(2022b)]%
        {Brodley22CACM}
\bibfield{author}{\bibinfo{person}{Carla~E. Brodley}.} \bibinfo{year}{2022}\natexlab{b}.
\newblock \showarticletitle{Why universities must resist GPA-based enrollment caps in the case of surging enrollments}.
\newblock \bibinfo{journal}{\emph{Commun. ACM}} \bibinfo{volume}{65}, \bibinfo{number}{8} (\bibinfo{date}{Aug.} \bibinfo{year}{2022}), \bibinfo{pages}{20--22}.
\newblock
\urldef\tempurl%
\url{https://doi.org/10.1145/3544547}
\showDOI{\tempurl}


\bibitem[Brodley and Gill(2024)]%
        {Brodley24CACM}
\bibfield{author}{\bibinfo{person}{Carla~E. Brodley} {and} \bibinfo{person}{Catherine Gill}.} \bibinfo{year}{2024}\natexlab{}.
\newblock \showarticletitle{The {B}{P}{C} relevance of common assessment in the introductory sequence}.
\newblock \bibinfo{journal}{\emph{Commun. ACM}} (\bibinfo{date}{to appear} \bibinfo{year}{2024}).
\newblock


\bibitem[Bruce et~al\mbox{.}(2003)]%
        {bruce03}
\bibfield{author}{\bibinfo{person}{Kim~B. Bruce}, \bibinfo{person}{Robert L.~Scot Drysdale}, \bibinfo{person}{Charles Kelemen}, {and} \bibinfo{person}{Allen Tucker}.} \bibinfo{year}{2003}\natexlab{}.
\newblock \showarticletitle{Why math?}
\newblock \bibinfo{journal}{\emph{Commun. ACM}} \bibinfo{volume}{46}, \bibinfo{number}{9} (\bibinfo{date}{sep} \bibinfo{year}{2003}), \bibinfo{pages}{40–44}.
\newblock
\showISSN{0001-0782}
\urldef\tempurl%
\url{https://doi.org/10.1145/903893.903918}
\showDOI{\tempurl}


\bibitem[Cheema and Galluzzo(2013)]%
        {cheema2013analyzing}
\bibfield{author}{\bibinfo{person}{Jehanzeb~R. Cheema} {and} \bibinfo{person}{Gary Galluzzo}.} \bibinfo{year}{2013}\natexlab{}.
\newblock \showarticletitle{Analyzing the gender gap in math achievement: Evidence from a large-scale US sample}.
\newblock \bibinfo{journal}{\emph{Research in Education}} \bibinfo{volume}{90}, \bibinfo{number}{1} (\bibinfo{year}{2013}), \bibinfo{pages}{98--112}.
\newblock


\bibitem[Chen et~al\mbox{.}(2020)]%
        {chen2020high}
\bibfield{author}{\bibinfo{person}{Chen Chen}, \bibinfo{person}{Jane~M. Kang}, \bibinfo{person}{Gerhard Sonnert}, {and} \bibinfo{person}{Philip~M. Sadler}.} \bibinfo{year}{2020}\natexlab{}.
\newblock \showarticletitle{High school calculus and computer science course taking as predictors of success in introductory college computer science}.
\newblock \bibinfo{journal}{\emph{ACM Transactions on Computing Education (TOCE)}} \bibinfo{volume}{21}, \bibinfo{number}{1} (\bibinfo{year}{2020}), \bibinfo{pages}{1--21}.
\newblock


\bibitem[De~Mol(2015)]%
        {de2015some}
\bibfield{author}{\bibinfo{person}{Liesbeth De~Mol}.} \bibinfo{year}{2015}\natexlab{}.
\newblock \showarticletitle{Some reflections on mathematics and its relation to computer science}.
\newblock \bibinfo{journal}{\emph{Automata, Universality, Computation: Tribute to Maurice Margenstern}} (\bibinfo{year}{2015}), \bibinfo{pages}{75--101}.
\newblock


\bibitem[Dougherty(2017)]%
        {dougherty2017math}
\bibfield{author}{\bibinfo{person}{John~P. Dougherty}.} \bibinfo{year}{2017}\natexlab{}.
\newblock \showarticletitle{MATH COUNTS Does mathematics serve computing as a support or a barrier?}
\newblock \bibinfo{journal}{\emph{ACM Inroads}} \bibinfo{volume}{8}, \bibinfo{number}{1} (\bibinfo{year}{2017}), \bibinfo{pages}{31--32}.
\newblock


\bibitem[Ellis et~al\mbox{.}(2016)]%
        {Ellis16}
\bibfield{author}{\bibinfo{person}{Jessica Ellis}, \bibinfo{person}{Bailey~K. Fosdick}, {and} \bibinfo{person}{Chris Rasmussen}.} \bibinfo{year}{2016}\natexlab{}.
\newblock \showarticletitle{Women 1.5 times more likely to leave STEM pipeline after calculus compared to men: Lack of mathematical confidence a potential culprit}.
\newblock \bibinfo{journal}{\emph{PLOS ONE}} \bibinfo{volume}{11}, \bibinfo{number}{7} (\bibinfo{date}{07} \bibinfo{year}{2016}), \bibinfo{pages}{1--14}.
\newblock
\urldef\tempurl%
\url{https://doi.org/10.1371/journal.pone.0157447}
\showDOI{\tempurl}


\bibitem[Florida\:Shines(2023)]%
        {excesshours}
\bibfield{author}{\bibinfo{person}{Florida\:Shines}.} \bibinfo{year}{2023}\natexlab{}.
\newblock \bibinfo{title}{Surcharge for Excess Hours}.
\newblock
\newblock
\urldef\tempurl%
\url{https://www.floridashines.org/excess-credit-surcharge}
\showURL{%
\tempurl}


\bibitem[Joint\:Task\:Force\:on\:Computing\:Curricula\:Association\:for\:Computing\:Machinery\:(ACM)\:and\:IEEE\:Computer\:Society(2013)]%
        {ACM2013}
\bibfield{author}{\bibinfo{person}{Joint\:Task\:Force\:on\:Computing\:Curricula\:Association\:for\:Computing\:Machinery\:(ACM)\:and\:IEEE\:Computer\:Society}.} \bibinfo{year}{2013}\natexlab{}.
\newblock \showarticletitle{Computer Science Curricula 2013: Curriculum Guidelines for Undergraduate Degree Programs in Computer Science}.
\newblock  (\bibinfo{year}{2013}).
\newblock
\showISBNx{9781450323093}


\bibitem[Konvalina et~al\mbox{.}(1983)]%
        {konvalina1983math}
\bibfield{author}{\bibinfo{person}{John Konvalina}, \bibinfo{person}{Stanley~A. Wileman}, {and} \bibinfo{person}{Larry~J. Stephens}.} \bibinfo{year}{1983}\natexlab{}.
\newblock \showarticletitle{Math proficiency: A key to success for computer science students}.
\newblock \bibinfo{journal}{\emph{Commun. ACM}} \bibinfo{volume}{26}, \bibinfo{number}{5} (\bibinfo{year}{1983}), \bibinfo{pages}{377--382}.
\newblock


\bibitem[Kumar et~al\mbox{.}(2023)]%
        {ACM2023gamma}
\bibfield{author}{\bibinfo{person}{Amruth~N. Kumar}, \bibinfo{person}{Monica~D. Anderson}, \bibinfo{person}{Brett~A. Becker}, \bibinfo{person}{Richard~L. Blumenthal}, \bibinfo{person}{Michael Goldweber}, \bibinfo{person}{Pankaj Jalote}, \bibinfo{person}{Susan Reiser}, \bibinfo{person}{Titus Winters}, \bibinfo{person}{Rajendra~K. Raj}, \bibinfo{person}{Sherif~G. Aly}, \bibinfo{person}{Douglas Lea}, \bibinfo{person}{Michael Oudshoorn}, \bibinfo{person}{Marcelo Pias}, \bibinfo{person}{Christian Servin}, \bibinfo{person}{Qiao Xiang}, \bibinfo{person}{Eric Eaton}, {and} \bibinfo{person}{Susan~L. Epstein}.} \bibinfo{year}{2023}\natexlab{}.
\newblock \bibinfo{title}{Computer Science Curricula 2023: Version Gamma}.  (\bibinfo{year}{2023}).
\newblock
\newblock
\shownote{Most recent draft of the 2023 ACM CS guidlines, published August}.


\bibitem[Lionelle et~al\mbox{.}(2024)]%
        {Lionelle24}
\bibfield{author}{\bibinfo{person}{Albert Lionelle}, \bibinfo{person}{Mc{K}enna Quam}, \bibinfo{person}{Carla~E.\ Brodley}, {and} \bibinfo{person}{Catherine Gill}.} \bibinfo{year}{2024}\natexlab{}.
\newblock \showarticletitle{Does curricular complexity in computer science influence the representation of women {C}{S} graduates?}. In \bibinfo{booktitle}{\emph{Proceedings of the 55th ACM Technical Symposium on Computing Science Education}} \emph{(\bibinfo{series}{SIGCSE '24}, Vol.~\bibinfo{volume}{1})}. \bibinfo{publisher}{ACM}, \bibinfo{address}{New York, NY}.
\newblock


\bibitem[Lockwood et~al\mbox{.}(2019)]%
        {LOCKWOOD2019100688}
\bibfield{author}{\bibinfo{person}{Elise Lockwood}, \bibinfo{person}{Anna~F. DeJarnette}, {and} \bibinfo{person}{Matthew Thomas}.} \bibinfo{year}{2019}\natexlab{}.
\newblock \showarticletitle{Computing as a mathematical disciplinary practice}.
\newblock \bibinfo{journal}{\emph{The Journal of Mathematical Behavior}}  \bibinfo{volume}{54} (\bibinfo{year}{2019}), \bibinfo{pages}{100688}.
\newblock
\showISSN{0732-3123}
\urldef\tempurl%
\url{https://doi.org/10.1016/j.jmathb.2019.01.004}
\showDOI{\tempurl}


\bibitem[Matti~Tedre and Malmi(2018)]%
        {histCS}
\bibfield{author}{\bibinfo{person}{Simon Matti~Tedre} {and} \bibinfo{person}{Lauri Malmi}.} \bibinfo{year}{2018}\natexlab{}.
\newblock \showarticletitle{Changing aims of computing education: A historical survey}.
\newblock \bibinfo{journal}{\emph{Computer Science Education}} \bibinfo{volume}{28}, \bibinfo{number}{2} (\bibinfo{year}{2018}), \bibinfo{pages}{158--186}.
\newblock
\urldef\tempurl%
\url{https://doi.org/10.1080/08993408.2018.1486624}
\showDOI{\tempurl}
\showeprint{https://doi.org/10.1080/08993408.2018.1486624}


\bibitem[Muzny et~al\mbox{.}(2024)]%
        {Muzny24}
\bibfield{author}{\bibinfo{person}{Felix Muzny}, \bibinfo{person}{Megan Giordano}, \bibinfo{person}{Emma Stone}, {and} \bibinfo{person}{Carla~E.\ Brodley}.} \bibinfo{year}{2024}\natexlab{}.
\newblock \showarticletitle{Collecting, analyzing, and acting on intersectional, longitudinal data and pass/fail/withdraw rates in computing courses}. In \bibinfo{booktitle}{\emph{Proceedings of the 55th ACM Technical Symposium on Computing Science Education}} \emph{(\bibinfo{series}{SIGCSE '24}, Vol.~\bibinfo{volume}{1})}. \bibinfo{publisher}{ACM}, \bibinfo{address}{New York, NY}.
\newblock


\bibitem[Muzny and Shah(2023)]%
        {Muzny23}
\bibfield{author}{\bibinfo{person}{Felix Muzny} {and} \bibinfo{person}{Michael~D. Shah}.} \bibinfo{year}{2023}\natexlab{}.
\newblock \showarticletitle{Teaching assistant training: An adjustable curriculum for computing disciplines}. In \bibinfo{booktitle}{\emph{Proceedings of the 54th ACM Technical Symposium on Computing Science Education}} \emph{(\bibinfo{series}{SIGCSE 2023}, Vol.~\bibinfo{volume}{1})}, \bibfield{editor}{\bibinfo{person}{Lina Battestilli}, \bibinfo{person}{Brian Dorn}, {and} \bibinfo{person}{Leen-Kiat Soh}} (Eds.). \bibinfo{publisher}{ACM}, \bibinfo{address}{New York, NY}.
\newblock


\bibitem[National\:Center\:for\:Education\:Statistics\:(NCES)(2013)]%
        {nontraditional}
\bibfield{author}{\bibinfo{person}{National\:Center\:for\:Education\:Statistics\:(NCES)}.} \bibinfo{year}{2013}\natexlab{}.
\newblock \bibinfo{title}{Nontraditional Undergraduates}.
\newblock
\newblock
\urldef\tempurl%
\url{https://nces.ed.gov/pubs2002/2002012.pdf}
\showURL{%
\tempurl}
\newblock
\shownote{(Retrieved December 2023)}.


\bibitem[National\:Center\:for\:Education\:Statistics\:(NCES)(2023)]%
        {IPEDS}
\bibfield{author}{\bibinfo{person}{National\:Center\:for\:Education\:Statistics\:(NCES)}.} \bibinfo{year}{2023}\natexlab{}.
\newblock \bibinfo{title}{The Integrated Postsecondary Education Data System}.
\newblock \bibinfo{howpublished}{\url{https://nces.ed.gov/ipeds/}}.
\newblock
\newblock
\shownote{(Retrieved July 2023)}.


\bibitem[Oudshoorn et~al\mbox{.}(2018)]%
        {Oudshoorn18}
\bibfield{author}{\bibinfo{person}{Michael~J. Oudshoorn}, \bibinfo{person}{Stan Thomas}, \bibinfo{person}{Rajendra~K. Raj}, {and} \bibinfo{person}{Allen Parrish}.} \bibinfo{year}{2018}\natexlab{}.
\newblock \showarticletitle{Understanding the new ABET Computer Science Criteria}. In \bibinfo{booktitle}{\emph{Proceedings of the 49th ACM Technical Symposium on Computer Science Education}} (Baltimore, Maryland, USA) \emph{(\bibinfo{series}{SIGCSE '18})}. \bibinfo{publisher}{Association for Computing Machinery}, \bibinfo{address}{New York, NY, USA}, \bibinfo{pages}{429–434}.
\newblock
\showISBNx{9781450351034}
\urldef\tempurl%
\url{https://doi.org/10.1145/3159450.3159534}
\showDOI{\tempurl}


\bibitem[Ralston(1981)]%
        {ralston1981computer}
\bibfield{author}{\bibinfo{person}{Anthony Ralston}.} \bibinfo{year}{1981}\natexlab{}.
\newblock \showarticletitle{Computer science, mathematics, and the undergraduate curricula in both}.
\newblock \bibinfo{journal}{\emph{The American Mathematical Monthly}} \bibinfo{volume}{88}, \bibinfo{number}{7} (\bibinfo{year}{1981}), \bibinfo{pages}{472--485}.
\newblock


\bibitem[Ralston(2005)]%
        {ralston2005we}
\bibfield{author}{\bibinfo{person}{Anthony Ralston}.} \bibinfo{year}{2005}\natexlab{}.
\newblock \showarticletitle{Do we need any mathematics in computer science curricula?}
\newblock \bibinfo{journal}{\emph{ACM SIGCSE Bulletin}} \bibinfo{volume}{37}, \bibinfo{number}{2} (\bibinfo{year}{2005}), \bibinfo{pages}{6--9}.
\newblock


\bibitem[Ross et~al\mbox{.}(2022)]%
        {ross2022removing}
\bibfield{author}{\bibinfo{person}{Monique Ross}, \bibinfo{person}{Mark~A. Weiss}, \bibinfo{person}{Lilia Minaya}, \bibinfo{person}{Andrew Laginess}, \bibinfo{person}{Disha Patel}, {and} \bibinfo{person}{Kathleen Quardokus~Fisher}.} \bibinfo{year}{2022}\natexlab{}.
\newblock \showarticletitle{Removing a barrier: Analysis of the impact of removing calculus and physics from CS on employability, salary, and broadening participation}. In \bibinfo{booktitle}{\emph{Proceedings of the 53rd ACM Technical Symposium on Computer Science Education-Volume 1}}. \bibinfo{pages}{460--466}.
\newblock


\bibitem[Tims et~al\mbox{.}(2023)]%
        {tims2023-inroads}
\bibfield{author}{\bibinfo{person}{Jodi~L. Tims}, \bibinfo{person}{Cindy Tucker}, \bibinfo{person}{Mark~A. Weiss}, {and} \bibinfo{person}{Stuart Zweben}.} \bibinfo{year}{2023}\natexlab{}.
\newblock \showarticletitle{Computing enrollment and retention: Results from the 2021--22 undergraduate enrollment cohort}.
\newblock \bibinfo{journal}{\emph{ACM Inroads}} \bibinfo{volume}{14}, \bibinfo{number}{4} (\bibinfo{date}{Nov} \bibinfo{year}{2023}), \bibinfo{pages}{24–43}.
\newblock
\showISSN{2153-2184}
\urldef\tempurl%
\url{https://doi.org/10.1145/3629980}
\showDOI{\tempurl}


\bibitem[Tucker et~al\mbox{.}(2001)]%
        {tucker2001our}
\bibfield{author}{\bibinfo{person}{Allen~B. Tucker}, \bibinfo{person}{Charles~F. Kelemen}, {and} \bibinfo{person}{Kim~B. Bruce}.} \bibinfo{year}{2001}\natexlab{}.
\newblock \showarticletitle{Our curriculum has become math-phobic!}. In \bibinfo{booktitle}{\emph{Proceedings of the thirty-second SIGCSE technical symposium on Computer Science Education}}. \bibinfo{pages}{243--247}.
\newblock


\bibitem[Ventura~Jr(2005)]%
        {ventura05}
\bibfield{author}{\bibinfo{person}{Philip~R. Ventura~Jr}.} \bibinfo{year}{2005}\natexlab{}.
\newblock \showarticletitle{Identifying predictors of success for an objects-first CS1}.
\newblock \bibinfo{journal}{\emph{Computer Science Education}}  \bibinfo{volume}{15} (\bibinfo{date}{09} \bibinfo{year}{2005}), \bibinfo{pages}{223--243}.
\newblock
\urldef\tempurl%
\url{https://doi.org/10.1080/08993400500224419}
\showDOI{\tempurl}


\bibitem[Wilson and Shrock(2001)]%
        {wilson2001contributing}
\bibfield{author}{\bibinfo{person}{Brenda~Cantwell Wilson} {and} \bibinfo{person}{Sharon Shrock}.} \bibinfo{year}{2001}\natexlab{}.
\newblock \showarticletitle{Contributing to success in an introductory computer science course: A study of twelve factors}.
\newblock \bibinfo{journal}{\emph{ACM SIGCSE Bulletin}} \bibinfo{volume}{33}, \bibinfo{number}{1} (\bibinfo{year}{2001}), \bibinfo{pages}{184--188}.
\newblock


\bibitem[Zhu et~al\mbox{.}(2024)]%
        {ross2024}
\bibfield{author}{\bibinfo{person}{Jia Zhu}, \bibinfo{person}{Disha Patel}, \bibinfo{person}{Camila Olivero-Araya}, \bibinfo{person}{Albert Lluberes}, \bibinfo{person}{Monique Ross}, \bibinfo{person}{Mark Weiss}, {and} \bibinfo{person}{Kathleen~Quardokus Fisher}.} \bibinfo{year}{2024}\natexlab{}.
\newblock \bibinfo{title}{A Mixed-Method Study on Undergraduate Students' Rational Choice between a Bachelor of Science or Bachelor of Arts in Computer Science (manuscript in preparation)}.  (\bibinfo{year}{2024}).
\newblock


\end{thebibliography}
